# Paper-based Flexible Supercapacitors with drawn van der Waals materials

Bahare Nouri[1], Andres Castellanos-Gomez[2], Foad Ghasemi[1]*

[1]Nanoscale Physics Device Lab (NPDL), Department of Physics, University of Kurdistan, 66177-15175, Sanandaj, Iran.

[2]Materials Science Factory, Instituto de Ciencia de Materiales de Madrid (ICMM-CSIC), E-28049 Madrid, Spain.

*Corresponding author: f.ghasemi@uok.ac.ir

**Abstract:**

Two-dimensional (2D) materials are widely used in various applications due to their extraordinary properties. In particular, their electrochemical stability, low electrical resistance, and huge specific surface area make them very interesting active materials for supercapacitors. Herein, flexible, biodegradable, and low-cost supercapacitors are introduced in a very simple way based on hand-drawing pencil traces or -rubbing molybdenum disulfide ($MoS_2$), titanium trisulfide ($TiS_3$) and franckeite traces on the paper. Results demonstrate that pencil-drawn paper has higher capacitance performance (~6.39 F/g) among the suggested electrodes. Interestingly, the introduced $MoS_2$/pencil, $TiS_3$/pencil, and franckeite/pencil drawn paper electrodes reveal dramatic improvements with long cyclic life thanks to the occurrence of synergetic effects and higher available active cites within the heterostructures. Moreover, the assembled symmetric solid-state supercapacitors retain their performance even under applied bending, indicating their excellent potential for wearable/flexible applications.

## 1. Introduction

Sustainable global economic growth requires more energy generation/consumption[1]. Therefore, energy storage devices are in high demand among consumers, making their development a crucial priority[2]. As a storage device, researchers have been interested in supercapacitors due to their high-power density, long-term cycling, and fast charge/discharge rates[3]. Moreover, with the development of wearable and internet of things (IoT) sensors, biodegradable flexible supercapacitors are more demanded than ever since conventional IoT devices constituent a big electronic-waste problem[4]. As a result, low-cost and high-performance flexible supercapacitors are required that not only be environmentally friendly but also do not disturb the main function of the sensors[5]. Up to now, various materials have been introduced as active layers in supercapacitors, such as carbon derivatives[6], metal oxides[7], metal hydroxides[8], polymers[9], etc. In addition to these materials, two-dimensional (2D) materials are also widely used in supercapacitors due to their environmental stability, enormous surface area, and ease of charge exchange[10]. Nevertheless, the poor capacitive performance of 2D materials is one of the major challenges that need to be addressed by combining them with other nanostructures[11]. Despite this, 2D materials offer high elasticity and remarkable in-plane mechanical properties that make them attractive for flexible energy storage devices[12, 13]. Many flexible supercapacitors have been so far proposed based on 2d materials such as rGO/$Ti_3C_2T_x$[14], $Ti_3C_2$/$WS_2$[15], graphene/CNT[16], 3D carbon/$MoS_2$[17], and $Ti_3C_2$/$MoS_2$/$Cu_2O$[18] that have shown excellent performance.

One of the flexible, cheap, lightweight, and biodegradable substrates is paper, which is suitable for wearable and skin-contact applications[19]. Recently, various paper-based supercapacitors are introduced, such as rGO/AgNPs[20], $MoS_2$/$SnS_2$ QDs[21], $MoS_2$/PANI/CNT[22],

MnO$_2$/PANI[23], Graphene/Bismuth ferrite[24], Graphite/CuNPs[25], and so on. In most of these reports, the paper surface is either modified with active materials or covered by electrochemical deposition methods[20]. However, the fabrication of prototype electrodes is time-consuming, complex, and in some cases expensive[26-28]. Instead, using commercial pencil traces on paper is a very simple, fast, and green method that can provide high-quality prototype electrodes on any type of paper substrate. Commercial pencils are classified under different labels from H to B, with the H labels having harder leads and a larger percentage of clay, while the other type has a softer lead and contains a larger content of graphite[29]. Hence, most pencil-drawn paper reports have employed B-type pencils, which provide better conductivity and surface coverage[30-32]. For example, Down et al. showed that handwriting B-type pencils on paper resulted in a capacitance value of about 101.6 µF/mg[33]. Furthermore, the hand-drawing method can be applied to other materials - notably 2d materials that are highly capable in electrochemical and energy storage applications - which can effectively enhance the performance of paper-based supercapacitors[32, 34].

In this report, pencil traces are employed for supercapacitor fabrications along with MoS$_2$, TiS$_3$, and franckeite traces on the paper substrate. The method of electrode fabrication is based on hand-drawing a 4B commercial pencil or -rubbing van der Waals bulk crystal on paper, which represents an easy, fast, low-cost, and environmentally friendly method. The electrochemical testing of the electrodes shows that the pencil-drawn electrode has the highest supercapacitor performance among all electrodes. By combining pencil traces with other layered materials, it is observed that the capacitance of the electrodes increases by several hundred percent compared to individual electrodes thanks to the effective role of the pencil trace in creating a uniform conductive platform and the occurrence of synergistic effects. Furthermore, the capacitive behavior of the assembled solid-state supercapacitor (i.e., TiS$_3$/pencil) demonstrates a negligible difference in the flat state and under bending, which indicates its excellent flexible performance.

## 2. Experimental

*2.1. Material*

Bulk franckeite crystal was acquired in mineral form from the San Jose mine in Bolivia. Molybdenum disulfide (MoS$_2$) crystal was obtained in mineral form from the Wolfram mine in Australia. Titanium trisulfide (TiS$_3$) was synthesized (see next subsection). Titanium powder was bought from Goodfellow. N-Methyl-2-pyrrolidone (NMP) solvent, sulfuric acid (H$_2$SO$_4$, 98%), polyvinyl alcohol (PVA, MW 72000), and sulfur powder were purchased from Merck. Ordinary kraft, tracing, and filter papers were supplied from local markets as substrates. 4B pencils were purchased from different brands of Sketch, sky stationery, Thermo bond, and Raffine sketch.

*2.2. Growth of TiS$_3$ microflakes*

TiS$_3$ microflakes were synthesized by the titanium sulfurization method as follows: the mixture of Ti and sulfur powders with a molar ratio of 1:3 was sealed in a vacuum ampoule and exposed to a temperature of 500 °C for 24 hours. After cooling to room temperature, the TiS$_3$ microflakes were collected from the ampoule.

*2.3. Electrode fabrications*

First, the papers were cut with dimensions of 1×1 cm$^2$. Then the pencil lead (or MoS$_2$, TiS$_3$, or franckeite crystals) was mechanically rubbed on the surface of the paper to completely cover it. The samples were placed on a wooden tongue depressor and double-sided copper tape was contacted to a part of the paper surface containing the film. Finally, epoxy resin was used to passivate the surface of the copper tape. To fabricate the hybrid electrodes, first, the paper surface was completely covered with pencil traces and then MoS$_2$ (TiS$_3$ or franckeite) crystal was rubbed on the surface containing the pencil traces.

*2.4. Preparation of PVA/H$_2$SO$_4$ electrolyte gel*

2 g PVA was slowly added to 10 ml deionized water while stirring. Then, the temperature of the solution was increased to 85 °C and 2 ml of sulfuric acid (H$_2$SO$_4$, 98%) was added dropwise to the solution. The solution was stirred for 3 h at 85 °C until a transparent gel was obtained.

*2.5. Electrochemical measurements*

A three-electrode electrochemical system containing platinum, Ag/AgCl, and samples were used as a counter, reference, and working electrodes, respectively. 1 M sodium sulfate (Na$_2$SO$_4$) solution was used as an electrolyte. Scan rates of 5 to 200 mV/s were used for CV tests. The GCD curves were obtained at different current densities from 0.1 to 0.5 A/g.

*2.6. Material Characterization*

Scanning electron microscopy (FE-SEM, TESCAN MIRA III LMH) with energy-dispersive X-ray Spectroscopy was used to investigate the morphology and composition of electrodes. The Ocean Insight, QEPRO-FL spectrometer was employed to obtain Raman spectra. The Panalytical X`Pert Pro X-ray diffractometer was utilized to obtain the XRD spectrum of the grown TiS$_3$ with CuKα radiation.

## 3. Result and discussions:

In the first step, the pencil trace is hand-drawn on three different types of kraft, filter, and tracing papers. Because the goal is to fabricate a supercapacitor with the best performance, electrochemical tests are performed on all three samples. **Fig. 1(a)** compares the CV curves of all three samples in 1 M Na$_2$SO$_4$ solution at a scan rate of 100 mV/s. Accordingly, the pencil-drawn tracing paper shows a higher current density compared with the other two samples. Due to its flat structure, it allows a uniform deposition of pencil traces on it. In contrast, it seems that achieving a uniform layer in the other two papers faces some difficulty due to the many gaps between the fibers, which leads to a difference in electrical resistance across the papers. Moreover, the high physical strength of the tracing paper in the ionic solution results in its better performance than the other two samples. **Fig. 1(b)** compares the gravimetric capacitance of all three samples. The method of calculating the capacitance is explained in the next section. According to it, the pencil-drawn tracing paper has a capacitance of about 1.59 F/g at a scan rate of 100 mV/s, which is more than pencil-drawn filter (0.98 F/g) and kraft (0.24 F/g) papers. Therefore, tracing paper was chosen as the suitable substrate for further tests. Panel (**a**) and (**b**) of **Fig. S1** show the SEM and its higher-magnified images of the relatively smooth surface of the tracing paper, respectively. EDX analysis proves the presence of carbon, oxygen, calcium, sodium, and chlorine elements in the paper (**Fig. S1(c)**). According to the elemental mapping in **Fig. S1(d)**, carbon and oxygen elements are uniformly distributed over the paper. **Table S1** provides the percentage composition of the tracing paper elements.

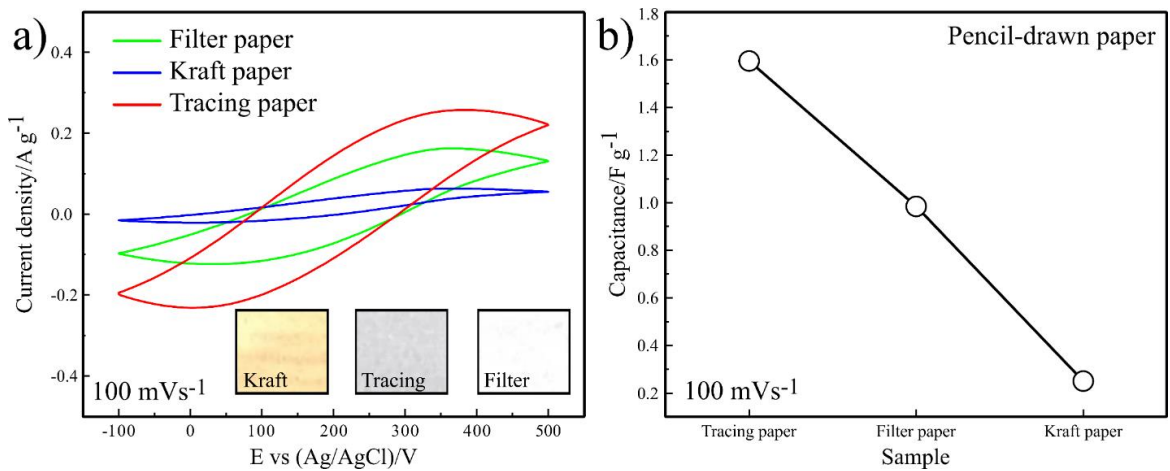

**Fig. 1.** (**a**) CV curves of the pencil-drawn tracing, filter, and kraft papers at a scan rate of 100 mV/s. The i**nset** shows the photograph of the raw papers. (**b**) Gravimetric capacitance of the pencil-drawn tracing, filter, and kraft papers.

In the following step, different 4B commercial pencil traces are investigated on tracing paper to determine their electrical resistance. In type B (2B, 4B, 6B, 8B) pencils, with an increasing number, pencil lead is softer and has more graphite percentage[29]. Therefore, it is expected that the B family covers the surface of the paper better than the H family which has more clay contents with harder lead[29]. In addition, there are several reports that 4B pencils provide lower electrical resistance compared to other B types[30-32]. Herein, different types of 4B commercial pencils are chosen to measure their electrical resistance as a function of trace numbers. According to **Fig. S2**, the Sky stationery commercial pencil shows the lowest resistance, and its value decreases with increasing the number of traces. With this kind of commercial pencil, it seems that the paper surface is effectively covered. **Fig. 2(a)** shows the SEM image of the pencil traces covering the surface of the tracing paper. Its elemental mapping indicates that carbon is uniformly distributed across the paper (**Fig. 2(b)**). The EDX spectrum of pencil traces on tracing paper is also presented in **Fig. S3**. According to it, carbon, oxygen, and silicon are the dominant elements, respectively. **Table S2** lists the percentage of contributing elements in the sample. By comparing the results of pencil-drawn paper with bare paper, it is found that in the former, the atomic percentage of carbon increases from 58.25% to 86.67%, while the atomic percentage of oxygen decreases from 41.56% to 9.42%. These results indicate that the carbon is well-deposited on the tracing paper. In addition to pencil-, $MoS_2$-, $TiS_3$- and franckeite-drawn papers are prepared by the same method. In the case of $MoS_2$, its bulk crystal was rubbed on the surface of the paper. **Fig. 2(c)** shows the SEM image of the $MoS_2$ film on paper where its elemental mapping reports the four dominant elements of carbon, oxygen, molybdenum, and sulfur (**Fig. 2(d)**). **Table S3** shows the element contents of the sample. According to the methods section, $TiS_3$ microcrystals are obtained by sulfurization of Ti powder at a temperature of 500 °C[34]. Based on the SEM image of the grown microcrystals presented in **Fig. S4(a)**, they have a length of several hundred and a width of a few microns. The XRD spectrum confirms the crystalline structure of $TiS_3$ microcrystals (**Fig. S4(b)**). Also, according to EDX analysis shown in **Table S4**, titanium and sulfur elements have 67.05 and 32.95 weight percentages with a stoichiometric ratio of 3.04. After the preparation of the $TiS_3$, its grown microflakes were rubbed on the paper. The SEM image of the sample (**Fig. 2(e)**) indicates that TiS$_3$ film covers the surface of the paper more uniformly. Moreover, the mapping analysis (**Fig. 2(f)**) shows the presence of the titanium and sulfur elements across the paper surface. **Table S5** provides the percentage of its elements in the sample. To fabricate the franckeite sample, its bulk crystal was rubbed on paper same as the two previous samples. According to the SEM image in **Fig. 2(g)**, franckeite flakes with smaller lateral sizes cover the surface of the paper. The elemental mapping shows the presence of carbon, oxygen, lead, and tin elements in the sample (**Fig. 2(h)**), whose percentages are presented in **Table S6**.

Raman spectra of the pencil-, TiS$_3$-, MoS$_2$- and franckeite-drawn papers are presented in panels **(i)** to **(l)** of **Fig. 2**. For pencil traces, three peaks are observed at 1305.75, 1582.42, and 2703.42 cm$^{-1}$, which are related to D, G, and 2D oscillation modes, respectively[35]. TiS$_3$ film has four peaks at 175.11, 305.18, 385.12, and 570.31 cm$^{-1}$, corresponding to the A$_g$ mode[34]. In the case of MoS$_2$, the peaks at 379.25 and 570.31 cm$^{-1}$ attribute to E$^1_{2g}$ and A$_{1g}$ modes, respectively[36]. Finally, the Raman spectrum of franckeite includes consecutive peaks at 66.11, 145.15, 194.01, 253.42, and 318.52 cm$^{-1}$ with a broad shoulder found in the range of 400 to 700 cm$^{-1}$, which are characteristic peaks of the franckeite's flakes reported in the literature[37].

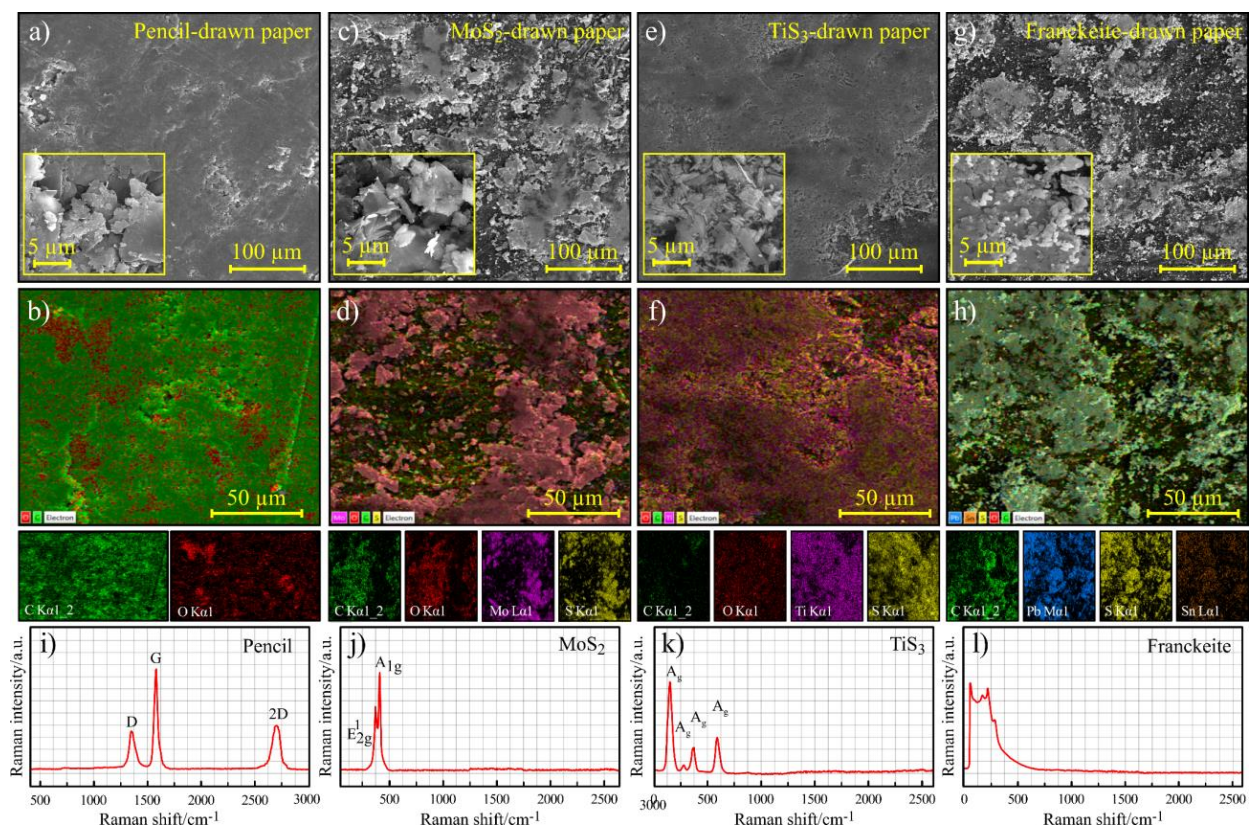

**Fig. 2.** SEM image of (**a**) pencil-, (**c**) MoS$_2$-, (**e**) TiS$_3$-, and (**g**) franckeite-drawn papers. **Insets** show the higher-magnified images. EDX mapping analysis of (**b**) pencil-, (**d**) MoS$_2$-, (**f**) TiS$_3$-, and (**h**) franckeite-drawn papers with their corresponding dominant elements. Raman spectra of (**i**) pencil, (**j**) MoS$_2$, (**k**) TiS$_3$ and (**l**) franckeite traces on the paper substrate.

The prepared samples are tested as working electrodes in a three-electrode electrochemical cell in 1 M Na$_2$SO$_4$ solution. **Fig. 3(a-d)** show the results of the measured CV curves at scan rates of 5 to 200 mV/s. Almost all the curves show a quasi-rectangular behavior indicating their acceptable capacitive performance[38]. **Fig. 3(a)** displays the CV curves of the pencil-drawn electrode in a potential window of -0.1 to 0.4 V. At the scan rate of 200 mV/s, the current density reaches more than 0.4 A/g, which refers to the suitable electrical conductivity of the film and its high potential in interacting with the electrolyte ions. **Fig. 3(b)** demonstrates the CV characteristics of the MoS$_2$-drawn electrode where the current density is appeared to be lower than that of the pencil electrode. It is primarily due to the non-uniformity of the film on the paper that results in a noticeable decrease in the film's conductivity. In **Fig. 3(c)**, the CV curves of the TiS$_3$-drawn paper show a relatively better quasi-rectangular behavior than the other layered materials corresponding to the higher capacitive performance[39]. Despite its higher current density than $MoS_2$ (even franckeite), it is still below pencil traces, which occurs as a result of the non-uniformity of the film thickness. The CV characteristics of the franckeite-drawn paper reveal that the shape of these curves is somewhat different from other samples probably due to the presence of pseudocapacitance behavior[40]. As can be seen in **Fig. 3(d)**, the current density has the lowest value in this sample compared to other electrodes. There can be several factors that reduce the effective interaction between the franckeite film and the electrolyte, including its low conductivity, small lateral size, and discontinuity of the deposited film[41].

The gravimetric capacitance of the introduced electrodes is calculated according to the following equation:

$$C = \frac{\int I(V)dV}{mv\Delta V} \qquad (1)$$

where I is current, m is the mass of the active material, ν is the scan rate, and ΔV is the potential window[36]. Accordingly, the capacitance is a function of the scan rate. For pencil-drawn paper, the capacitance values of 6.39, 5.00, 3.66, 2.59, and 1.78. 1.43 F/g is obtained at scan rates of 5, 10, 20, 50, 100, and 200 mV/s, respectively. In the case of $MoS_2$-drawn paper, the values of 1.95, 1.58, 1.14, 0.61, 0.43, and 0.29 F/g are measured at the scan rates of 5 to 200 mV/s, respectively. The gravimetric capacitance of $TiS_3$-drawn paper was calculated to be 1.25 F/g at a scan rate of 5 mV/s, but decreased to 0.35 F/g at a scan rate of 200 mV/s. Finally, the values of 0.60 and 0.14 F/g are obtained for the franckeite-drawn electrode at the scan rate of 5 and 200 mV/s, respectively. **Table S7** provides capacitance values of all electrodes at different scan rates. The results of the calculated capacitances as a function of the scan rates are also presented in **Fig. 3(e)** Based on it, the pencil-drawn electrode has the highest capacitance compared to other electrodes. After that, the value of capacitance decreases from $MoS_2$- to $TiS_3$- and franckeite-drawn electrodes, respectively. The capacitance of the pencil-drawn sample has a value of 6.39 F/g at a scan rate of 5 mV/s, which is 69.48%, 80.43%, and 90.61% more than $MoS_2$-, $TiS_3$-, and franckeite-drawn samples, respectively. Therefore, among these layered materials, the pencil trace exhibits better capacitive performance than other layered material traces due to its better paper coverage and higher effective interaction with the electrolyte. Moreover, it can be seen that with increasing the scan rates, the capacitance decreases in all samples. In fact, the transfer of ions on the surface of the active material decreases with increasing the scan rate which results in reducing the capacitance values[36].

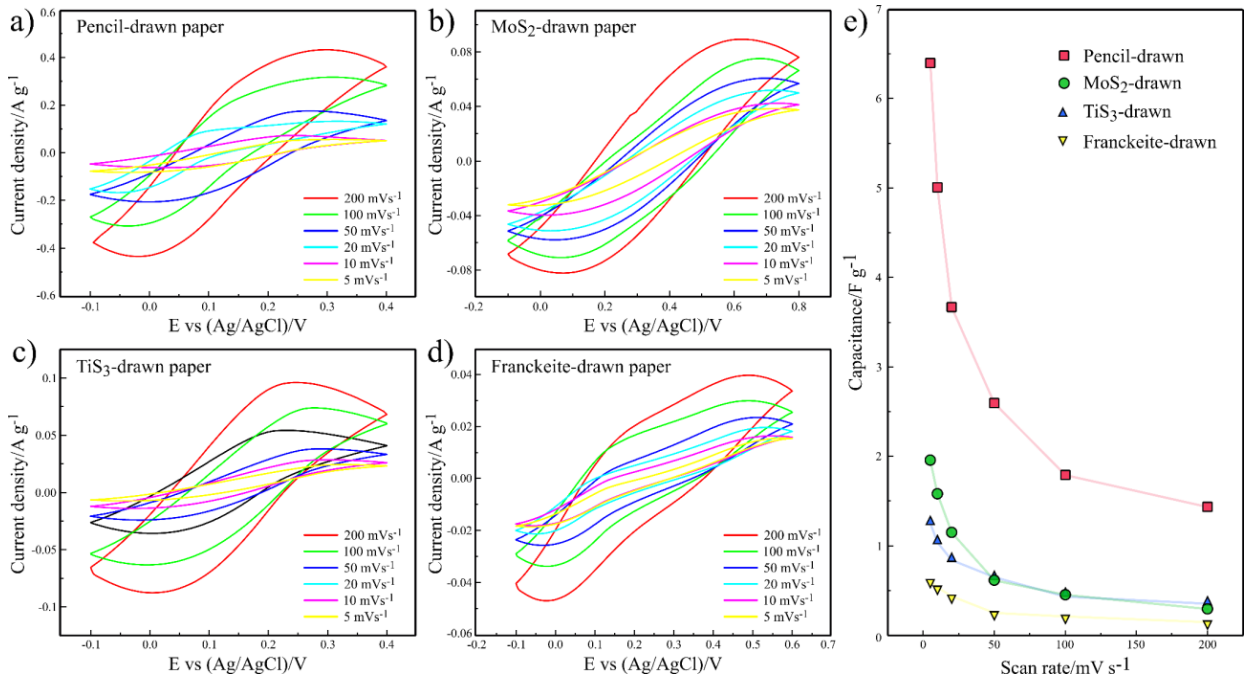

**Fig. 3.** The CV curves of (**a**) pencil-, (**b**) $MoS_2$-, (**c**) $TiS_3$-, (**d**) franckeite-drawn papers, and (**e**) gravimetric capacitance of the prepared electrodes at scan rates of 5 to 200 mV/s.

According to the results, the pencil traces almost cover the paper surface due to its soft structure, which also ends in a higher capacitance value than other electrodes. Therefore, three hybrid electrodes are introduced in which the pencil traces act as the bottom platform, and another layered material structure was rubbed on them. **Fig. 4(a)** displays the SEM image of $MoS_2$/pencil-drawn paper, where MoS$_2$ flakes are located on the pencil traces. Its elemental mapping (**Fig. 4(b)**) indicates that the dominant elements of carbon, oxygen, molybdenum, and sulfur are distributed throughout the surface. **Table S8** presents the abundance of composing elements of the sample. In **Fig. 4(c)**, an SEM image of the TiS$_3$/pencil sample can be seen, where TiS$_3$ is on top of the pencil traces. Indeed, both structures have appeared on the surface because the rubbing process ends with the approximate combination of both materials. The elemental mapping of the layer shows that carbon, oxygen, titanium, and sulfur are distributed over the surface (**Fig. 4(d)**). A list of the elements in the sample can be found in **Table S9**. The SEM image of the franckeite/pencil sample exhibits the presence of franckeite flakes on the pencil film (**Fig. 4(e)**), whose elemental mapping (**Fig. 4(f)**) confirms the presence of carbon, lead, sulfur, and tin elements. **Table S10** also displays the percentage of sample elements in detail.

The Raman spectra of all three introduced samples are provided in **Fig. 4(g-h)**. In all three samples, the D, G, and 2D characteristic peaks of the pencil traces are clearly visible[35]. In the MoS$_2$/pencil sample, in addition to these peaks, two $E^1_{2g}$ and $A_{1g}$ peaks are also presented[42]. For the TiS$_3$/pencil sample, all four $A_g$ peaks of the TiS$_3$ structure have appeared, and in the franckeite/pencil sample, franckeite peaks can be seen in the corresponding spectrum[34, 37]. As a result of Raman analysis, the hybrid samples appear to be well-formed[42].

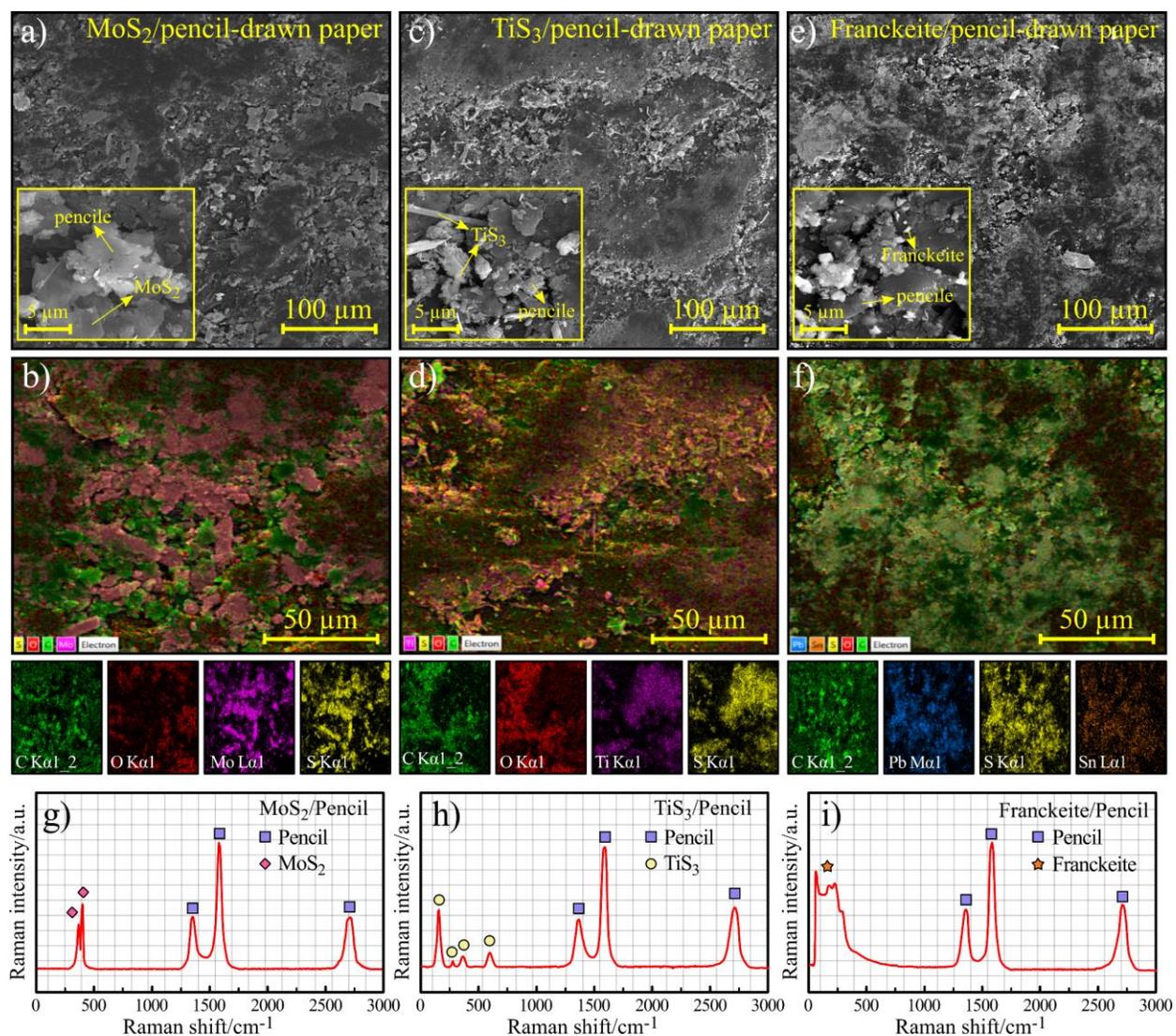

**Fig. 4.** SEM image of (**a**) MoS$_2$/pencil-, (**c**) TiS$_3$/pencil-, (**e**) franckeite/pencil-drawn papers. The higher-magnified images are presented in the insets. EDX mapping analysis of (**b**) MoS$_2$/pencil-, (**d**) TiS$_3$/pencil-, and (**f**) franckeite/pencil-drawn papers with their corresponding dominant elements. Raman spectra of the (**g**) MoS$_2$/pencil-, (**h**) TiS$_3$/pencil-, and (**i**) franckeite/pencil-drawn papers.

In the next step, the suggested samples are employed as a working electrode in a three-electrode electrochemical cell. The same potential window is considered for all electrodes since pencil traces serve as a bottom platform in all of them, and in this condition their performance can be comparable. **Fig. 5(a)** shows the CV curves of the MoS$_2$/pencil sample at 5 to 200 mV/s scan rates. It can be seen that the current density reaches more than 1 A/g, which is more than the previous individual samples. Moreover, the shape of the curves is almost quasi-rectangular, which shows its proper capacitive behavior[36]. CV curves of the TiS$_3$/pencil sample are presented in **Fig. 5(b)**. The current density is almost 2 A/g in this sample at a scan rate of 200 mV/s, which is higher than the MoS$_2$/pencil case. **Fig. 5(c)** also demonstrates the CV curves of the franckeite/pencil sample, which has a quasi-rectangular shape. However, their internal area seems smaller than the previous two samples. The gravimetric capacitance of each electrode was calculated according to equation (1) at different scan rates. The obtained results are plotted in **Fig. 5(d)**. Accordingly, the MoS$_2$/pencil has a capacitance of 14.84 F/g at a scan rate of 5 mV/s, which decreases to a value of 3.63 F/g at 200 mV/s. Increasing the scan rate from 5 to 200 mV/s leads to a decrease in capacitance value for TiS$_3$/pencil from 17.31 to 4.17 F/g. In franckeite/pencil, 8.67 F/g is obtained at a scan rate of 5 mV/s, while 1.70 F/g is calculated at a scan rate of 200 mV/s. **Table S11** shows the values of the capacitances at different scan rates. Accordingly, the capacitance of MoS$_2$/pencil, TiS$_3$/pencil, and franckeite/pencil have increased by 661%, 1248%, and 1345% compared to the individual MoS$_2$, TiS$_3$, and franckeite electrodes, respectively. This significant increase in capacitance can be due to two reasons: 1) the dominating role of pencil traces in providing a uniform conductive platform, and 2) the presence of synergistic effects between pencil traces and other layered structures. Apparently, this synergistic effect is more significant in TiS$_3$ and franckeite than MoS$_2$, since the highest capacitance improvement is observed in the first two cases.

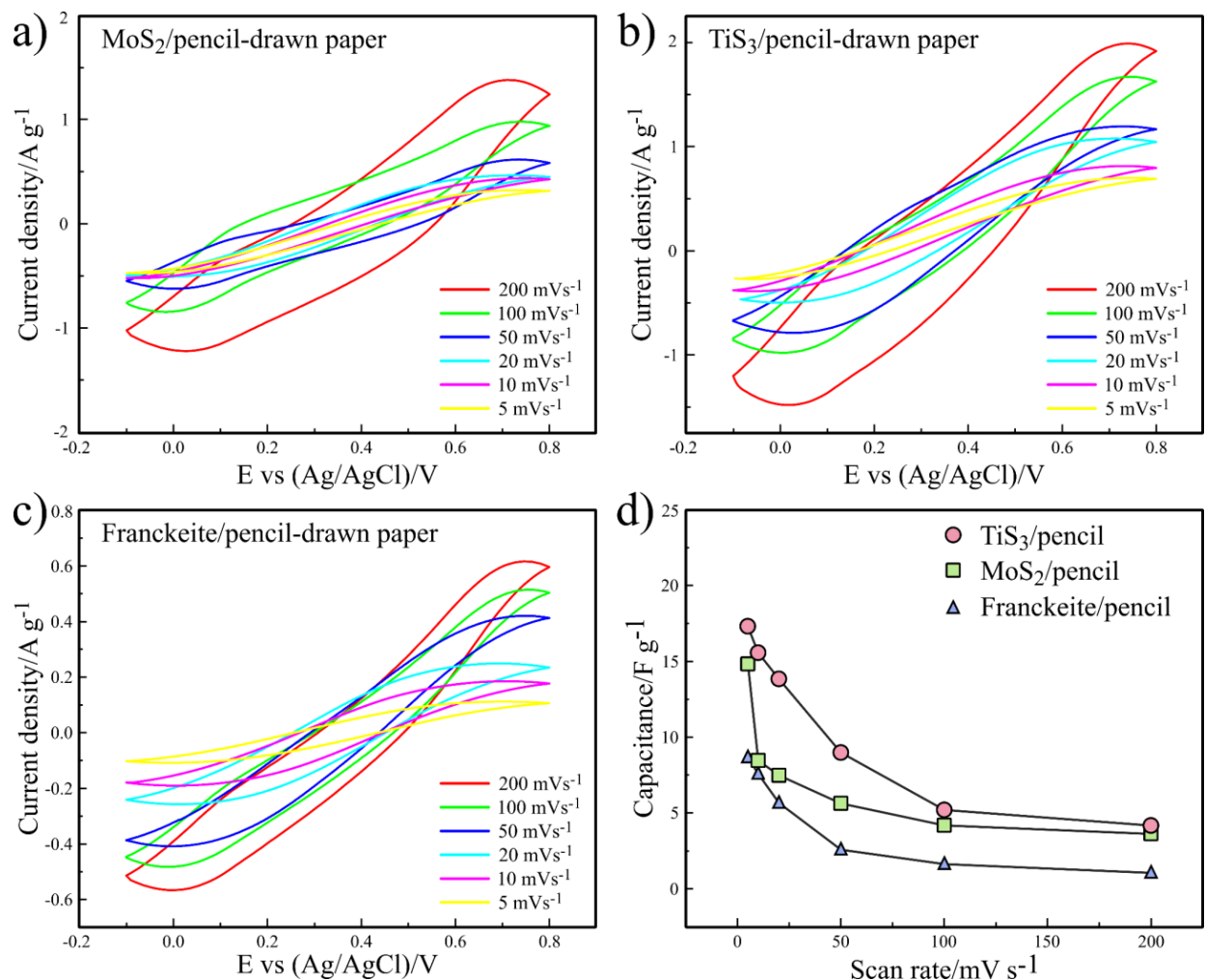

**Fig. 5.** CV Curves of (**a**) MoS$_2$/pencil-, (**b**) TiS$_3$/pencil-, and (**c**) franckeite/pencil-drawn paper electrodes at different scan rates of 5 to 200 mV/s. (**d**) Capacitance versus scan rates of the corresponding electrodes.

As can be seen in **Fig. 6(a)**, the charge-discharge test of all three introduced electrodes is compared at the same current density of 0.1 A/g. Accordingly, the TiS$_3$/pencil sample has a longer charge-discharge cycle compared with the other two electrodes. In detail, the discharge times of MoS$_2$/pencil, TiS$_3$/pencil, and franckeite/pencil were calculated to be ~153, ~164, and ~99 s, respectively. The gravimetric capacitance of the electrodes is also compared in **Fig. 6(b)** at various current densities of 0.1 to 0.5 A/g. Capacitance was calculated based on the following equation:

$$C = \frac{It}{m\Delta V} \qquad (2)$$

where I, t, m, and ΔV are the current, discharge time, the mass of active material, and potential window, respectively[36]. In all current densities, the capacitance of TiS$_3$/pencil is higher than the other two samples. In this sample, the capacitance changes from 18.2 to 11.6 F/g by increasing the current density from 0.1 to 0.5 A/g. The discharge time is also reduced from 164 s to 20.8 s. In the MoS$_2$/pencil, with increasing current density, the capacitance and discharge time decrease from 17.0 F/g and 153 s to 10.3 F/g and 18.5 s, respectively. In the case of franckeite/pencil, the capacitance and discharge time were decreased to 6.1 F/g and 10.9 s. In all three electrodes, a significant increase is observed in capacitance and discharge time compared to the pencil electrode with 7.3 F/g and 65.7 s at a current density of 0.1 A/g, respectively. **Table S12** and **S13** show the calculated capacitance and discharge time values of the hybrid and pencil electrode at different current densities respectively.

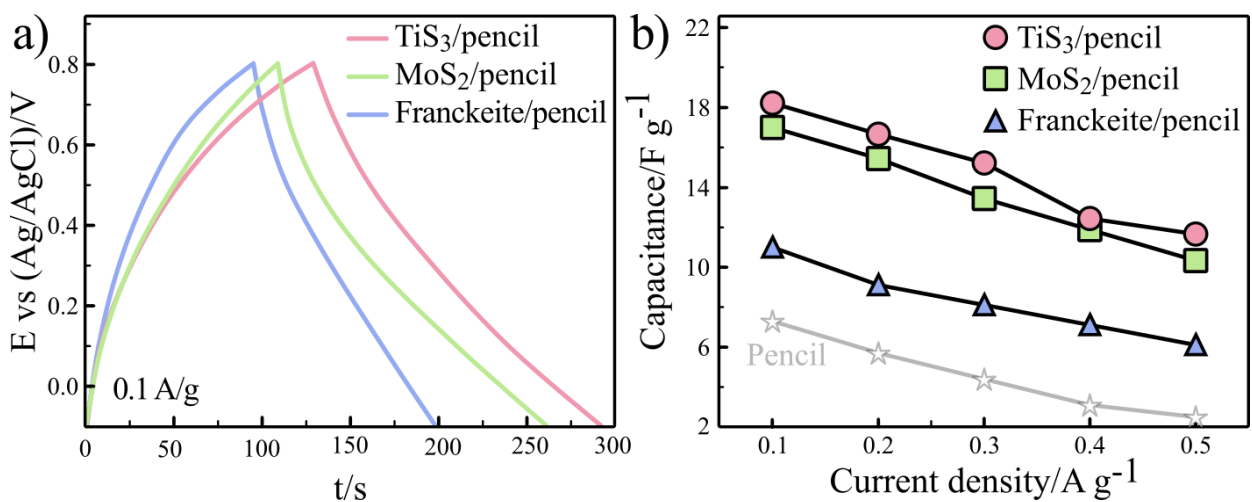

**Fig. 6.** (**a**) Charge-discharge curves of the MoS$_2$/pencil, TiS$_3$/pencil, and franckeite/pencil at a current density of 0.1 A/g. (**b**) Capacitance of the corresponding electrodes versus current densities of 0.1 to 0.5 A/g and comparison with the pencil electrode.

The cyclic stability of the introduced electrodes is studied for 1000 consecutive charge-discharge cycles at a scan rate of 200 mV/s. **Fig. 7(a)** shows that the capacitance of MoS$_2$/pencil retains 79.02% of its initial value after 1000 cycles. In the TiS$_3$/pencil sample, 89.06% of the initial capacitance remains after 1000 cycles (**Fig. 7(b)**). In the case of franckeite/pencil, the capacitance of the 1000th cycle is 90.75% of the first cycle capacitance (**Fig. 7(c)**). The insets in each Fig. show the CV curves of the first and last cycles. Regarding the point that the substrate is paper, all three electrodes demonstrate significant cyclic stability, which potentially enables them as practical energy storage devices. Moreover, the comparison of the obtained results with the pencil electrode shows their acceptable performance because the pencil electrode maintains 91.81% of its initial capacity after 1000 cycles according to **Fig. S5**. Based on this, the pencil trace plays a significant role in maintaining the cyclic life of the electrodes.

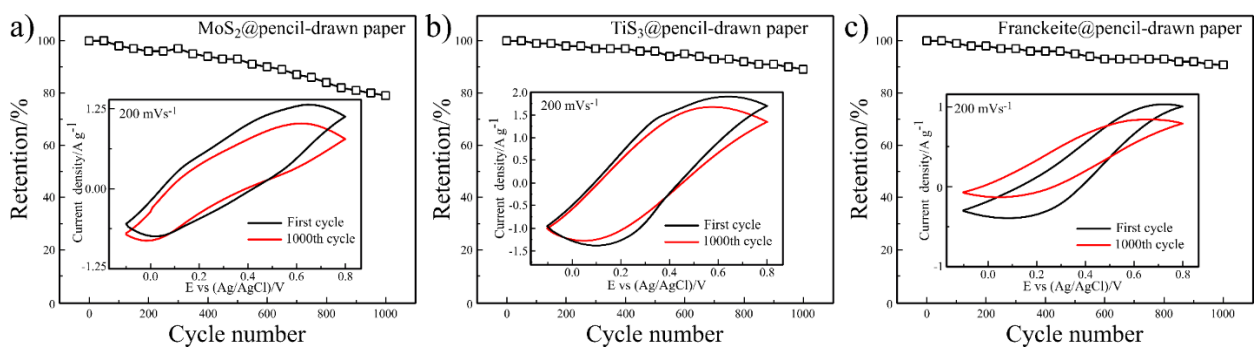

**Fig. 7.** Cyclic stability of the (**a**) MoS$_2$/pencil-, (**b**) TiS$_3$/pencil-, and (**c**) franckeite/pencil-drawn papers for 1000 consecutive charge-discharge cycles. **Insets** show the CV curves of the first and 1000$^{th}$ cycles.

The flexible performance of electrodes is investigated using a symmetric solid-state supercapacitor based on TiS$_3$/pencil-drawn paper. **Fig. 8(a)** shows the schematic of the electrode fabrication steps, as well as the photograph of the final device. For this purpose, an interdigital pattern is hand-drawn with pencil traces followed by rubbing TiS$_3$ on them. Next, PVA/H$_2$SO$_4$ gel is dropped on the active area of the pattern, and copper tape is attached as the contact. Then, the capacitive performance of the solid-state device is investigated in a potential window of -0.1 to 0.8 V vs Ag/AgCl at the scan rate of 100 mV/s in two states of flat and under applied bending. According to **Fig. 8(b)**, the CV curves of the device do not change appreciably even after bending. The result shows the excellent performance of the introduced device which make it as a very promising supercapacitor for wearable applications.

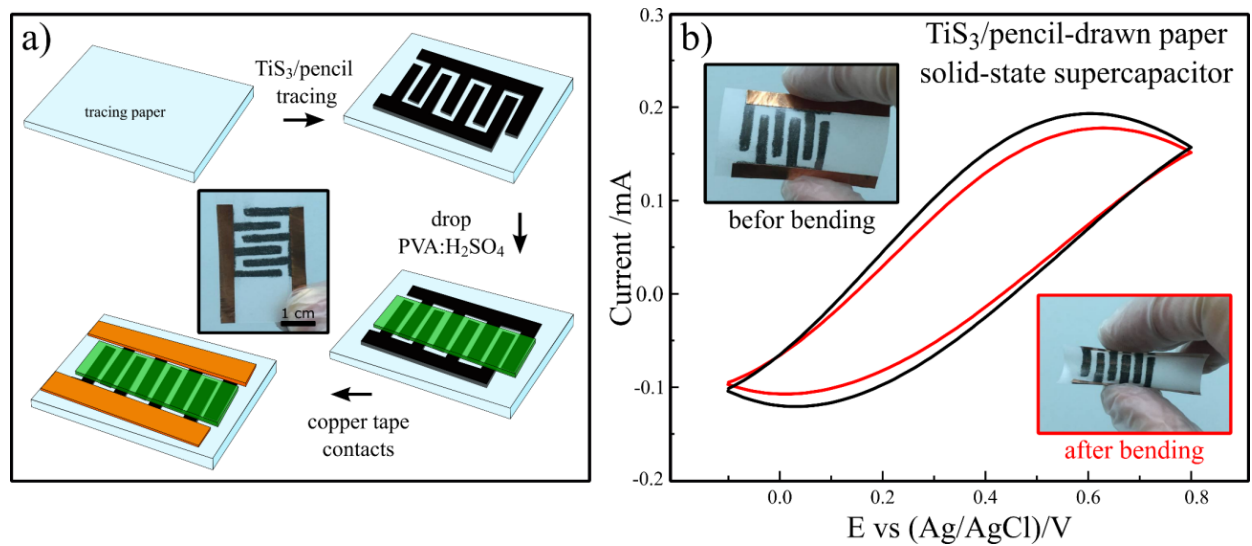

**Fig. 8.** (**a**) Schematic illustration of the fabrication steps of a symmetric solid-state supercapacitor based on the TiS$_3$/pencil active material. (**b**) CV curves of the solid-state TiS$_3$/pencil drawn paper supercapacitor before and after bindings at a scan rate of 100 mv/s.

## 4. Conclusion

2D materials are interesting in electrochemical applications due to their comparable surface area, high charge-carrier transport, and chemical stability. In order to increase their capacitance, various composite structures have been reported so far based on them. In this work, a simple, cheap, fast, and green method is introduced to fabricate flexible van der Waals-drawn supercapacitors on paper, which show excellent performances. Moreover, combining pencil traces with other layered materials of MoS$_2$, TiS$_3$, and franckeite leads to 661%, 1284 %, and 1345 % improvements in capacitances, respectively. The introduced symmetric solid-state supercapacitor also performs well under bending, which suggests their promising potential as flexible energy storage devices.

## References

[1] S.B. Selçuklu, M.D. Rodgers, A. Movlyanov, Computers & Industrial Engineering, 164 (2022) 107892.
[2] Z. Zhu, T. Jiang, M. Ali, Y. Meng, Y. Jin, Y. Cui, W. Chen, Chemical Reviews, 122 (2022) 16610-16751.
[3] Y. Zhao, Q. Liang, S.M. Mugo, L. An, Q. Zhang, Y. Lu, Advanced Science, 9 (2022) 2201039.

[4] J.T. Carvalho, I. Cunha, J. Coelho, E. Fortunato, R. Martins, L. Pereira, ACS Applied Energy Materials, 5 (2022) 11987-11996.
[5] T. Xu, H. Du, H. Liu, W. Liu, X. Zhang, C. Si, P. Liu, K. Zhang, Advanced Materials, 33 (2021) 2101368.
[6] P. Xie, W. Yuan, X. Liu, Y. Peng, Y. Yin, Y. Li, Z. Wu, Energy Storage Materials, 36 (2021) 56-76.
[7] C. An, Y. Zhang, H. Guo, Y. Wang, Nanoscale Advances, 1 (2019) 4644-4658.
[8] T. Nguyen, M.d.F. Montemor, Advanced Science, 6 (2019) 1801797.
[9] Z. Wang, M. Zhu, Z. Pei, Q. Xue, H. Li, Y. Huang, C. Zhi, Materials Science and Engineering: R: Reports, 139 (2020) 100520.
[10] F.R. Fan, R. Wang, H. Zhang, W. Wu, Chemical Society Reviews, 50 (2021) 10983-11031.
[11] K. Khan, A.K. Tareen, M. Aslam, A. Mahmood, Q. khan, Y. Zhang, Z. Ouyang, Z. Guo, H. Zhang, Progress in Solid State Chemistry, 58 (2020) 100254.
[12] W. Yu, K. Gong, Y. Li, B. Ding, L. Li, Y. Xu, R. Wang, L. Li, G. Zhang, S. Lin, Small, 18 (2022) 2105383.
[13] D. Zhou, L. Zhao, B. Li, Journal of Energy Chemistry, 62 (2021) 27-42.
[14] Y. Luo, X. Yin, Y. Luo, H. Xie, X. Bin, Y. Tian, M. Ju, W. Que, Advanced Materials Interfaces, 9 (2022) 2101619.
[15] J.V. Vaghasiya, C.C. Mayorga-Martinez, J. Vyskočil, Z. Sofer, M. Pumera, Advanced Functional Materials, 30 (2020) 2003673.
[16] X. Zang, Y. Jiang, M. Sanghadasa, L. Lin, Sensors and Actuators A: Physical, 304 (2020) 111886.
[17] Z. Jiang, S. Zhai, M. Huang, P. Songsiriritthigul, S.H. Aung, T.Z. Oo, M. Luo, F. Chen, Energy, 227 (2021) 120419.
[18] X. Mao, Y. Zou, F. Xu, L. Sun, H. Chu, H. Zhang, J. Zhang, C. Xiang, ACS Applied Materials & Interfaces, 13 (2021) 22664-22675.
[19] E.O. Polat, Advanced Materials Technologies, 6 (2021) 2000853.
[20] Q. Huang, Y. Yang, R. Chen, X. Wang, EcoMat, 3 (2021) e12076.
[21] P.K. Enaganti, V. Selamneni, P. Sahatiya, S. Goel, New Journal of Chemistry, 45 (2021) 8516-8526.
[22] I.W.P. Chen, Y.-C. Chou, P.-Y. Wang, The Journal of Physical Chemistry C, 123 (2019) 17864-17872.
[23] X. Ling, G. Zhang, Z. Long, X. Lu, Z. He, J. Li, Y. Wang, D. Zhang, Journal of Industrial and Engineering Chemistry, 99 (2021) 317-325.
[24] A. Soam, R. Kumar, M. C, M. Singh, D. Thatoi, R.O. Dusane, Journal of Alloys and Compounds, 813 (2020) 152145.
[25] A. Dutta, R. Nayak, M. Selvakumar, D. Devadiga, P. Selvaraj, S.S. Kumar, Materials Letters, 295 (2021) 129849.
[26] Y. Chen, Z. Yin, D. Huang, L. Lei, S. Chen, M. Yan, L. Du, R. Xiao, M. Cheng, Journal of Colloid and Interface Science, 611 (2022) 356-365.
[27] P.M. Pataniya, S. Dabhi, V. Patel, C.K. Sumesh, Surfaces and Interfaces, 34 (2022) 102318.
[28] N. Song, Y. Wu, W. Wang, D. Xiao, H. Tan, Y. Zhao, Materials Research Bulletin, 111 (2019) 267-276.
[29] V.N. Ataide, I.V.S. Arantes, L.F. Mendes, D.S. Rocha, T.A. Baldo, W.K.T. Coltro, T.R.L.C. Paixão, Journal of The Electrochemical Society, 169 (2022) 047524.
[30] S.H. Lee, J.Y. Ban, C.-H. Oh, H.-K. Park, S. Choi, Scientific Reports, 6 (2016) 28588.
[31] P. Rani, K.S. Kumar, A.D. Pathak, C.S. Sharma, Applied Surface Science, 533 (2020) 147483.
[32] W. Zhang, Q. Zhao, C. Munuera, M. Lee, E. Flores, J.E.F. Rodrigues, J.R. Ares, C. Sanchez, J. Gainza, H.S.J. van der Zant, J.A. Alonso, I.J. Ferrer, T. Wang, R. Frisenda, A. Castellanos-Gomez, Applied Materials Today, 23 (2021) 101012.
[33] M.P. Down, C.W. Foster, X. Ji, C.E. Banks, RSC Advances, 6 (2016) 81130-81141.
[34] E. Mahmoodi, M.H. Amiri, A. Salimi, R. Frisenda, E. Flores, J.R. Ares, I.J. Ferrer, A. Castellanos-Gomez, F. Ghasemi, Scientific Reports, 12 (2022) 12585.
[35] X. Liao, Q. Liao, X. Yan, Q. Liang, H. Si, M. Li, H. Wu, S. Cao, Y. Zhang, Advanced Functional Materials, 25 (2015) 2395-2401.

[36] M. Rashidi, F. Ghasemi, Electrochimica Acta, 435 (2022) 141379.
[37] A.J. Molina-Mendoza, E. Giovanelli, W.S. Paz, M.A. Niño, J.O. Island, C. Evangeli, L. Aballe, M. Foerster, H.S.J. van der Zant, G. Rubio-Bollinger, N. Agraït, J.J. Palacios, E.M. Pérez, A. Castellanos-Gomez, Nature Communications, 8 (2017) 14409.
[38] N. Namdar, F. Ghasemi, Z. Sanaee, Scientific Reports, 12 (2022) 4254.
[39] S. Amuthameena, K. Dhayalini, B. Balraj, C. Siva, N. Senthilkumar, Inorganic Chemistry Communications, 131 (2021) 108803.
[40] X. Mu, D. Wang, F. Du, G. Chen, C. Wang, Y. Wei, Y. Gogotsi, Y. Gao, Y. Dall'Agnese, Advanced Functional Materials, 29 (2019) 1902953.
[41] X. Gao, X. Du, T.S. Mathis, M. Zhang, X. Wang, J. Shui, Y. Gogotsi, M. Xu, Nature Communications, 11 (2020) 6160.
[42] F. Ghasemi, M. Hassanpour Amiri, Applied Surface Science, 570 (2021) 151228.

Graphical Abstract

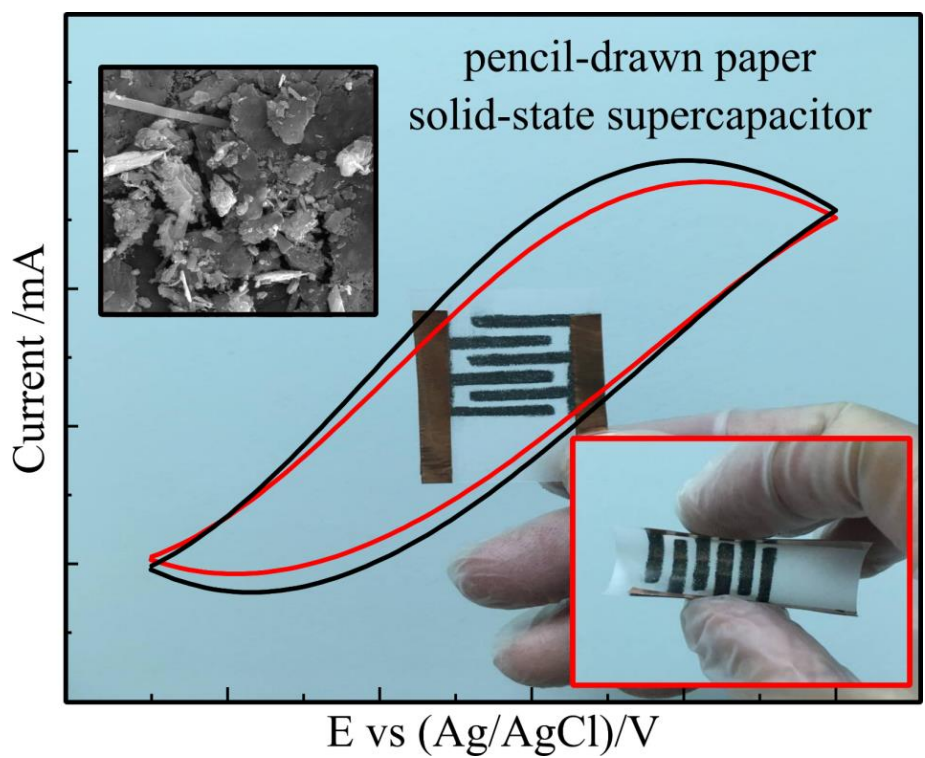